\documentstyle[12pt]{article}
\date{}
\def\be{\begin{equation}}
\def\ee{\end{equation}}
\def\bea{\begin{eqnarray}}
\def\eea{\end{eqnarray}}
\def\s{\sigma}
\def\al{\alpha}

\def\om{\omega}
\title{Special classes of solutions\\
 for linear string baryon configuration}
\author{ V.\,P. Petrov, G.\,S. Sharov\thanks{E-mail: german.sharov@tversu.ru}\\
{\small Tver state university}\\
{\small Tver, 170002, Sadovyj per. 35, Mathem. dep-t.}}
\begin{document}
\maketitle
\begin{abstract}
For the linear string baryon model with three material
points (three quarks) joined sequentially by the relativistic strings,
the class of motions admitting linearizable boundary conditions
is investigated.
These motions may be represented as the Fourier series with eigenfunctions
of some boundary-value problem.
The two types of rotational motions are found among the mentioned class
of solutions.
\end{abstract}

\bigskip
\noindent{\bf Introduction}
\medskip

Four various string models of baryon were suggested by X.~Artru \cite{AY}
They differ from each other by geometric character of junction
of three massive points (quarks) by relativistic strings.
Four variants are possible: a) the ``three-string" model
or Y configuration with three strings from three quarks joined
in the fourth massless point \cite{Collins,PY}; b) the ``triangle" model
or $\Delta$-configuration with pairwise connection of three quarks
by three relativistic strings \cite{Tr,PRTr}; c) the quark-diquark
model $q$-$qq$ \cite{Ko} (from the point of view of classical dynamics it
coincides with the meson model of relativistic string with massive
ends \cite{Ch,BN}); d) the linear configuration $q$-$q$-$q$ with
quarks connected in series \cite{lin,4B} (see Fig.~1).

\begin{figure}[hb]
\unitlength=2.0mm
\begin{center}
\begin{picture}(38,7)
\thicklines
\put(5,5){\line(1,0){28}}
\put(5,5){\circle*{1}}\put(19,5){\circle*{1}}\put(33,5){\circle*{1}}
\put(4.5,2){$q$}\put(18.5,2){$q$}\put(32.5,2){$q$}
\end{picture}\end{center}
\caption{Linear string baryon configuration $q$-$q$-$q$.}
\end{figure}
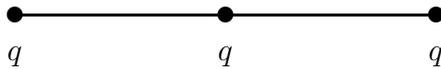

In the present work the latter model is considered.
It was not studied quantitativelly before Ref.~\cite{lin} where we
solved the initial-boundary value problem for classical
motion of this configuration and investigated the stability problem
for the rotational motion of this system. It was shown that this
motion (a flat uniform rotation of the rectilinear
string with the middle quark at rest at a center of rotation
\cite{Ko,4B}) is unstable. This instability results in the complicated
motion of the middle material point (quark) with quasi-periodical varying
of the distance between the nearest two quarks \cite{lin}.
But the system $q$-$q$-$q$ is not transformed in quark-diquark ($q$-$qq$) one,
as was supposed formerly in Ref.~\cite{Ko}.

In this paper for the system $q$-$q$-$q$ we study the class of motions
admitting linearizable boundary conditions. The similar motions for the
string with massive ends were first considered in Refs.~\cite{B77,BN77}
and they were exhaustively classified in Ref.~\cite{PeSh}.

For the $q$-$q$-$q$ string baryon configuration after the review of its
classical dynamics in Sect.~1 we consider in Sects.~2 the classes of
motions with linearizable boundary conditions.

\bigskip
\noindent{\bf 1. Classical dynamics of the linear string baryon configuration}
\medskip

Let's consider an open relativistic string with the tension $\gamma$
carrying three pointlike masses $m_1$, $m_2$, $m_3$ (the masses $m_1$
and $m_3$ are plased at the ends of the string).
The action for this system is \cite{lin}
\be
 S[X^\mu]=-\int\limits_{\tau_1}^{\tau_2}\! d\tau\left\{\gamma\!
\int\limits_{\s_1(\tau)}^{\s_3(\tau)}\!\!
\left[\big(\dot X,X'\big)^2-\dot X^2X'{}^2\right]^{\frac12}\!d\s+\sum
_{i=1}^3m_i\sqrt{X_i^{*2}(\tau)}\right\}.
\label{S}\ee
Here $X^\mu(\tau,\s)$ are coordinates of a point of
the string in $D$-dimensional Minkowski space $R^{1,D-1}$
with signature $(+,-,-,\dots)$,
the speed of light $c=1$,
$\,(\tau,\s)\in\Omega=\Omega_1\cup\Omega_2$ (Fig.\,2),
$\big( a,b\big)=a^\mu b_\mu$ --- (pseudo)scalar product,
$\dot X^\mu=\partial_\tau X^\mu$, $X'{}^\mu=\partial_\s X^\mu$,
$X_i^{*\mu}(\tau)=\frac d{d\tau}X^\mu(\tau,\s_i(\tau))$; $\s_i(\tau)$
($i=1,2,3$) --- inner coordinates of world lines of pointlike masses
(quarks).

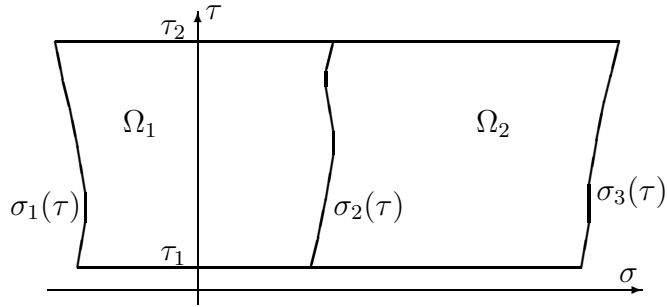
\begin{figure}[th]
\unitlength=1.0mm
\begin{center}
\begin{picture}(86,40)
\put(5,2){\vector(1,0){79}} \put(25,0){\vector(0,1){39}}
\thicklines
\put(9,5){\line(1,0){67}} \put(6,35){\line(1,0){75}}
\put(9,5){\line(1,6){1}} \put(10,11){\line(0,1){4}}
\put(10,15){\line(-1,6){1}} \put(9,21){\line(-1,5){1}}
\put(8,26){\line(-1,4){1}} \put(7,30){\line(-1,5){1}}
\put(40,5){\line(1,4){1}} \put(41,9){\line(1,5){1}}
\put(42,14){\line(1,6){1}} \put(43,20){\line(0,1){3}}
\put(43,23){\line(-1,6){1}} \put(42,29){\line(0,1){2}}
\put(42,31){\line(1,4){1}}
\put(76,5){\line(1,6){1}} \put(77,11){\line(0,1){5}}
\put(77,16){\line(1,6){1}} \put(78,22){\line(1,5){1}}
\put(79,27){\line(1,4){2}}
\put(81,3){$\sigma$} \put(26,38){$\tau$}
\put(20,36){$\tau_2$} \put(20,6){$\tau_1$}
\put(0,12){$\sigma_1(\tau)$} \put(43,12){$\sigma_2(\tau)$}
\put(78,14){$\sigma_3(\tau)$} \put(15,23){$\Omega_1$}
\put(62,23){$\Omega_2$}
\end{picture}
\caption{Domain of integration in Eq.~(1).}
\end{center}
\end{figure}

The equations of motion of the string and the boundary conditions
are derived form the action (\ref{S}) \cite{lin}. They have
the simplest form if with the help of nondegenerate reparametrization
$\tau=\tau(\tilde\tau,\tilde\s)$, $\s=\s(\tilde\tau,\tilde\s)$
the induced metric on the world surface of the string is made
continuous and conformally-flat \cite{Tr}, i.e., satisfies the orthonormality
conditions.
\be\dot X^2+X'{}^2=0,\qquad\big(\dot X,X'\big) = 0.
\label{ort}\ee

Under conditions (\ref{ort}) the equations of motion become linear
\be \ddot X^\mu-X''{}^\mu=0
\label{eq}\ee
and the boundary conditions take the simplest form
\bea
&m_i\frac d{d\tau}U^\mu_i(\tau)+\epsilon_i
\gamma\bigl[X'{}^\mu{}+\s_i'(\tau)\,\dot X^\mu\bigr]\bigg|_{\s=\s_i(\tau)}=0,
\quad i=1,3,&\label{qq}\\
&m_2\frac d{d\tau}U^\mu_2(\tau)-\gamma\big[X'{}^\mu+
\s_2'(\tau)\,\dot X^\mu\big]\bigg|_{\s=\s_2(\tau)+0}+
\gamma\big(X'{}^\mu+\s_2'(\tau)\,\dot X^\mu\big)
\bigg|_{\s=\s_2(\tau)-0}=0.&\label{qqq}
\eea
Here $\epsilon_1=-1$, $\epsilon_3=1$ and
$$
U^\mu_i(\tau)=\frac{X_i^{*\mu}(\tau)}{\sqrt{X_i^*{}^2(\tau)}}=
\frac{\dot X^\mu+\s_i'(\tau)\,X'{}^\mu}
{\sqrt{\dot X^2\cdot(1-\s_i'{}^2)}}\bigg|_{\s=\s_i(\tau)},\quad i=1,\,2,\,3
$$
are the unit $R^{1,D-1}$-velocity vector of $i$-th quark.

Derivatives of $X^\mu(\tau,\s)$ can have discontinuities
on the line $\s=\s_2(\tau)$.
However, the function $X^\mu(\tau,\s)$ and the tangential derivatives
$\frac d{d\tau}X^\mu(\tau,\s_2(\tau))$ are continuous.
Therefore the jumps of the functions
$\dot X^\mu$ and $X'{}^\mu$
are related by the condition
$$
[\dot X^\mu]+\s_2'(\tau)[X'{}^\mu]=0.
$$
Using this relation the boundary condition (\ref{qqq})
may be rewritten in the form
\be
m_2\frac d{d\tau}U^\mu_2(\tau){}+\gamma(1-\s_2'{}^2)
\bigl[X'{}^\mu(\tau,\s_2-0)-
X'{}^\mu(\tau,\s_2+0)\bigr]=0.\!\!
\label{qqq2}\ee

\bigskip
\noindent{\bf 1. Motions with linearizable boundary conditions}
\medskip

The present work is focused on world surfaces
supposing the parametrization satisfying the conditions (\ref{ort})
and also the following conditions:
\be
\sqrt{\dot X^2}\Big|_{\s=\s_i}=C_i=\mbox{const},\quad
\s_i(\tau)=\s_i={\mbox{const}},\quad i=1,2,3.
\label{7}\ee
To fulfil these conditions the parameter $\tau$ for the
quark trajectories is to be proportional to the natural parameter
$s=\int(\dot X^2)^{1/2}d\tau$ (the proper time).
Let's assume, without loss of generality, that
$\s_1=0$ and $\s_3=\pi$ \cite{lin}.
For the class of surfaces under consideration, constraint (\ref{7})
leads to linearization of boundary conditions (\ref{qq}), (\ref{qqq2}):
\be
\left.\bigl(\ddot X^\mu-Q_1X^{\prime\mu}\bigr)\right|_{\s=0}=0,\qquad
\left.\bigl(\ddot X^\mu+Q_3X^{\prime\mu}\bigr)\right|_{\s=\pi}=0.
\label{8}\ee
\be
\ddot X^\mu(\tau,\s_2)+Q_2\bigl[X'{}^\mu(\tau,\s_2-0)-
X'{}^\mu(\tau,\s_2+0)\bigr]=0.\!\label{9}\ee
Here the notation
$$Q_i=\gamma C_i/m_i,\qquad i=1,\,2,3;\qquad Q_i>0$$
is used.
The solutions to Eq.(\ref{eq}) with boundary conditions (\ref{8}), (\ref{9})
are constructed by the method of separation of variables(the Fourier method),
i. e., the solution is sought in the form of a linear combination
of the expressions
$X^\mu(\tau,\s)=\al^\mu u(\s)\,T(\tau)$, where $\al^\mu$ is an arbitrary
constant vector. The substitution of this expression into Eq.(\ref{eq})
results in the equations
\be
T''(\tau)+\om^2 T=0,\qquad u''(\s)+\om^2u=0.
\label{10}\ee
We write the nontrivial solution $u(\s)$ of Eq.(\ref{10}) for $\om\ne0$
in the form
\be
u(\s)=\left\{\begin{array}{ll}
A_1\cos\om\s+B_1\sin\om\s,& \s\in[\s_1,\s_2],\\
A_2\cos\om\s+B_2\sin\om\s,& \s\in[\s_2,\s_3].
\end{array}\right.
\label{11}\ee

From the physical point of view the function $u(\s)$ is continuous in the
segment $[\s_1,\s_3]$ but the $u'(\s)$ may have discontinuities for $\s=\s_2.$
The function $\alpha^\mu u(\s)\,e^{i\om\tau}$ is a solution of Eq.(\ref{eq}).
Boundary conditions (\ref{8}), (\ref{9}) impose constraints on
the $A_1,B_1,A_2,B_2.$

 The substitution of $X^\mu(\tau,\s)=\alpha^\mu u(\s)\,e^{i\om\tau}$
into conditions (\ref{8}), (\ref{9}) results to the system of equations
with respect to the $A_1,B_1,A_2,B_2.$
\be
\left\{\begin{array}{l}
\om A_1+Q_1B_1=0,\\
(\om C+Q_3S)\,A_2+(\om S-Q_3C)\,B_2=0,\\
(Q_2+\om C_2S_2Q_2)\,A_1+\om S_2^2B_1-Q_2A_2=0,\\
\om C_2^2A_1-(Q_2-\om C_2S_2)\,B_1+Q_2B_2=0.\\
\end{array}\right.
\label{12}\ee

Here the notations are used:
\,$\;C=\cos\pi\om$, $\;S=\sin\pi\om$, $\;C_2=\cos\om\s_2$,
$\;S_2=\sin\om\s_2$,
$\;C_*=\cos\om(\pi-\s_2)$, $\;S_*=\sin\om(\pi-\s_2)$.

System (\ref{12}) has nontrivial solution only if the following
condition is executed
\bea
&\om^3S_2S_*-\om^2(Q_1C_2S_*+Q_2S+Q_3S_2C_*)+{}&\nonumber\\
&{}+\om\big[Q_2(Q_1+Q_3)\,C+Q_1Q_3C_2C_*\big]+Q_1Q_2Q_3S=0.&
\label{13}\eea
This condition is equation with respect to the $\om$.

Transcendental Eq.(\ref{13}) has the countable set of real roots
and $\lim\limits_{n\to\infty}\om_n=+\infty$ if $\om=\om_n$ are numerated
in the order of increasing.

For every value of $\om=\om_n$ and an arbitrary value of $A_1$
we find $B_1,\,A_2,\,B_2$ from the system (\ref{12}).
In this case the function $\alpha^\mu u(\s)\,e^{i\om_n\tau}$ is the
solution of Eq.(\ref{eq}) and it also satisfies boundary conditions (\ref{8}),
(\ref{9}).

These facts let us search the solution of the problem (\ref{eq}),
(\ref{8}), (\ref{9}) in the form of a series
\be
X^\mu(\tau,\s)=x_0^\mu+p_0^\mu\tau+
\!\sum_{n\ne0}\!\alpha_n^\mu u_n(\s)\,e^{i\om_n\tau}.
\label{14}\ee
where
$\alpha_{-n}^\mu u_{-n}(\s)=\overline{\alpha_n^\mu u_n(\s)}$,
if $X^\mu$ is real.

Now we show that world surfaces of the form (\ref{14})
with the single frequency $\om\ne0$ is possible
\be
X^\mu(\tau,\s)=\al_0^\mu+p_0^\mu\tau+B_n\,u_n(\s)\,\bigl[e_n^\mu\cos\om_n\tau+
g_n^\mu\sin\om_n\tau\bigr].
\label{15}\ee
Let's require fulfillment of the condition: $u(\s)\in C^1(0,\pi).$
Taking into account (\ref{11}) that it condition is satisfied only
if $A_1=A_2=A$, $B_1=B_2=B$, we rewrite the solution (\ref{11}) in the form
$$
u(\s)=A\cos\om\s+B\sin\om\s,\qquad \s\in[\s_1,\s_3].
$$
The last two equations of system (\ref{12}) will be equivalent to one equation
\be
A\cos\om\s_2+B\sin\om\s_2=0.\label{16}\ee
The first two equations of system (\ref{12}) may be represented as follows

\be
\left\{\begin{array}{l}
\om A+Q_1B=0,\\
(\om C+Q_3S)\,A+(\om S-Q_3C)\,B=0.
\end{array}\right.\label{17}\ee
System (\ref{17}) has nontrivial solutions only if
 $\om$ is a root of the equation \cite{PeSh}

\be
(Q_1+Q_3)\,\om \cos\pi \om=(\om^2-Q_1Q_3)\sin\pi\om.\label{18}\ee
Transcendental  equation (\ref{18}) has the countable set of real roots
$\om=\om_n$, $n\in Z.$
If $\om=\om_n$ the solutions $u_n(\s)$ take the form
\be
u_n(\s)=A\Big(\cos\om_n\s-\frac{\om_n}{Q_1}\sin\om_n\s\Big),\quad\s\in[0,\pi].
\label{19}\ee
The function $u_n(\s)$ has $n$ roots on the interval $(0,\pi).$
Let's require that the value $\s_2$ coincides with any of these roots:
\be
\cos\om_n\s_2-\frac{\om_n}{Q_1}\sin\om_n\s_2=0.
\label{20}\ee
Equality (\ref{20}) for function (\ref{19}) results
in the fulfillment of condition (\ref{16}).
Therefore, function (\ref{19}) is the solution of system
(\ref{16}), (\ref{17}). Here $\om_n$ is the root of Eq. (\ref{18})
and $\s_2$ is the root of Eq.(\ref{20}).
It means that function (\ref{15}) satisfies to boundary conditions
(\ref{8})-(\ref{9}).

The substitution shows that function (\ref{15}) with $u_n(\s)$ in form
(\ref{19}) is the solution of Eq. (\ref{eq}) and satisfies to conditions
(\ref{ort}), (\ref{7}) if $p_0^2=(AB_n)^2\om_n^2(1+\om_n^2/Q_1^2)$ and
the vectors $p^\mu_0$, $e^\mu_n$ and $g^\mu_n$ are pairwise orthogonal and
$e_n^2=g^2_n=-1$.

The solution (\ref{15}), (\ref{19}) describes the uniform rotation
of the $n$ times folded rectilinear string in the plain with the vectors
$e^\mu_n$, $g^\mu_n$. The massive endpoints $m_1$, $m_3$ move along the
circles and the middle point is at the rotational center, i.~e. at the zero
point $\s=\s_2$ of the function $u_n(\s)$, satisfying Eq.~(\ref{20}).
The world surface of the solution (\ref{15}), (\ref{19}) is helicoid.

The uniform rotations of the folded rectilinear open (massless) string
were founded by Y.~Nambu \cite{Nambu}, and for the string with massive
ends they were classified in Ref.~\cite{PeSh}.

Note that in the particular case $\s_2=\pi/2$ the equation (\ref{13})
after  division by $S_2=S_*=\sin^2(\pi\om/2)$ may be solved with respect
to $\cot(\pi\om/2)$:
$$
\cot\frac{\pi\om}2=\frac{(Q_1+2Q_2+Q_3)\,\om^2-2Q_1Q_2Q_3\pm
\sqrt{D(\om)}}{2Q^2\om},
$$
where $D(\om)=[(Q_1-Q_3)^2+4Q_2^2]\,\om^4+4Q_2^2(Q_1^2+Q_3^2)\,\om^2+
4Q_1^2Q_2^2Q_3^2$.

And in the symmetrical case $m_1=m_3$, $Q_1=Q_3$,
$\s_2=\pi/2$ these two equations (or Eq.~(\ref{13})) take the form:
\be
\cot\frac{\pi\om}2=\frac\om{Q_1},\qquad
\cot\frac{\pi\om}2=\frac{\om^2-2Q_1Q_3}{(Q_1+2Q_2)\,\om}.
\label{symm}\ee
The roots of the first Eq.~(\ref{symm}) correspond to the solutions
(\ref{15}), (\ref{19}) with the middle quark at the rotational center.
But the roots of the second Eq.~(\ref{symm})  result in the rotations
of twice folded string with the quark $m_2$ at one end (at the bend point)
and two other massive points at the other end.

\bigskip
\noindent{\bf Conclusion}
\medskip

In the present work the we obtain the class of solutions for
the linear string baryon model $q$-$q$-$q$, generalizing the simple
rotational motion, that may be presented as (\ref{15}), (\ref{19}) with $n=1$.
In Ref.~\cite{lin} the stability of this solution was investigated
numerically and was shown that it is unstable in Lyapunov's sense ---
any small asymmetric perturbation grows and results in complicated motion
with quasi-periodical varying of the distance between
the nearest two quarks. However the minimal value of the mentioned distance
$\Delta R$ does not equal zero, in other words, the system $q$-$q$-$q$ is not
transformed in quark-diquark ($q$-$qq$) one, as was supposed in Ref.~\cite{Ko}.

The class of motions (\ref{15}), (\ref{19}) obtained in this work has
the same nature with the mentioned simple motion. So they are to be unstable
too, but the rotations corresponding the roots of the second Eq.~(\ref{symm})
are candidates to be unique stable solutions for this string configuration.

This analysis is important for choosing the most adequate string baryon
model and  for the progress in quantization of nonlinear string models
carrying masses on the basis of their quasirotational states \cite{Exc}.

\smallskip

The work is supported by the Russian Foundation of Basic Research,
grant  00-02-17359.

\end{document}